\documentclass[slac_one]{revtex4}
\usepackage{graphicx}
\usepackage{fancyhdr}
\usepackage{amsmath}
\newcommand{\Et}{\not\negthickspace{E}_{T}}
\newcommand{\ttbar}{t\bar{t}}

\begin{document}

\title{Top quark pair cross section measurement at ATLAS} 

\author{M. Gosselink (for the ATLAS collaboration)}
\affiliation{Nikhef, Amsterdam, The Netherlands}

\begin{abstract}
An accurate determination of the top quark pair production cross section at the LHC provides a valuable check of the Standard Model. Given the high statistics which will be available (about one top quark pair per second, at a luminosity of $10^{33} $cm$^{-2}$s$^{-1}$), this check can be performed relatively fast after the turn on of the LHC. The prospects for measuring the total top quark pair cross section with the ATLAS detector during the initial period of LHC running is presented here. The cross section is determined in the single lepton channel and in the dilepton channel. 
\end{abstract}

\maketitle

\thispagestyle{fancy}

\section{INTRODUCTION}
At the LHC, top quark pairs ($\ttbar$) will be produced mainly via gluon fusion \mbox{($\sim$ 90\%)}. A recent prediction for the cross section at $\sqrt{s}=14$ TeV is a next-to-leading order (NLO) calculation with soft-gluon next-to-leading-log (NLL) resummation \cite{xsecNLONLL}: $\sigma_{\ttbar}^{NLO+NLL}=908^{+82}_{-85}$ (factorisation and renormalisation scales) $^{+30}_{-29}$ (parton distribution function (PDF)) pb. As $V_{tb} \approx 1$, top quarks nearly always decay into a $W$-boson and a $b$-quark. A $W$-boson decays roughly 2/3$^{rd}$ of the cases into two quarks and 1/3$^{rd}$ into a lepton and a neutrino. Therefore, $\ttbar$ decay can be divided into three channels: 4/9$^{th}$ fully hadronic, 4/9$^{th}$ semileptonic and 1/9$^{th}$ dileptonic. The fully hadronic channel suffers from large QCD multi-jet background. The presence of missing transverse energy (from undetectable neutrino's) and at least one lepton in the two other channels creates the possibility to select $\ttbar$ events while reducing background considerably. These two channels are used in ATLAS for $\ttbar$ cross section measurements with the first 100pb$^{-1}$ integrated luminosity of data \citep{ATLAS-CSC} and will be discussed in the following sections.

\section{SINGLE LEPTON CHANNEL}
Two complementary measurements in the single lepton channel are investigated: a counting method and a likelihood fit. Events are selected that passed the electron (muon) trigger {\texttt{e22i}} ({\texttt{mu20}}) and contain missing transverse energy $\Et > 20$ GeV, an isolated electron (muon) with transverse momentum $p_{T} > 20$ GeV/c and pseudo rapidity $\lvert \eta \rvert < 2.5$, three jets with $p_{T} > 40$ GeV/c and $\lvert \eta \rvert < 2.5$ and an additional fourth jet with $p_{T} > 20$ GeV/c and $\lvert \eta \rvert < 2.5$. The hadronic top is reconstructed by taking the invariant mass $M_{jjj}$ of the three jet combination with the highest $p_{T}$. To reduce background from jet combinatorics and other processes, at least one di-jet combination is required to be compatible with the $W$-boson mass $\lvert M_{jj}-M_{W} \rvert < 10$ GeV/c$^{2}$. 
\begin{figure*}[ht]
\centering
\includegraphics[width=.3\textwidth]{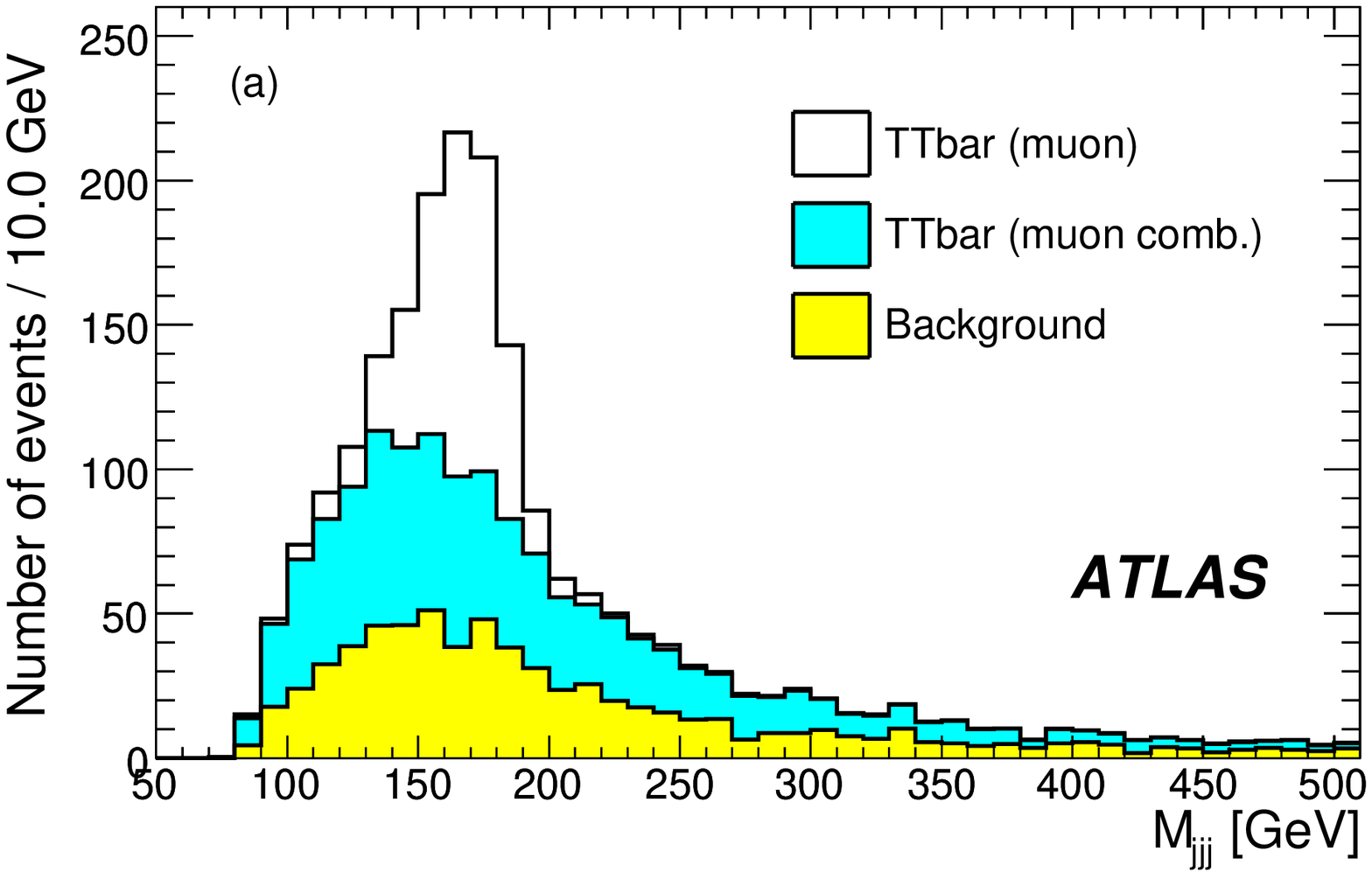}
\hspace{.1\textwidth}
\includegraphics[width=.3\textwidth]{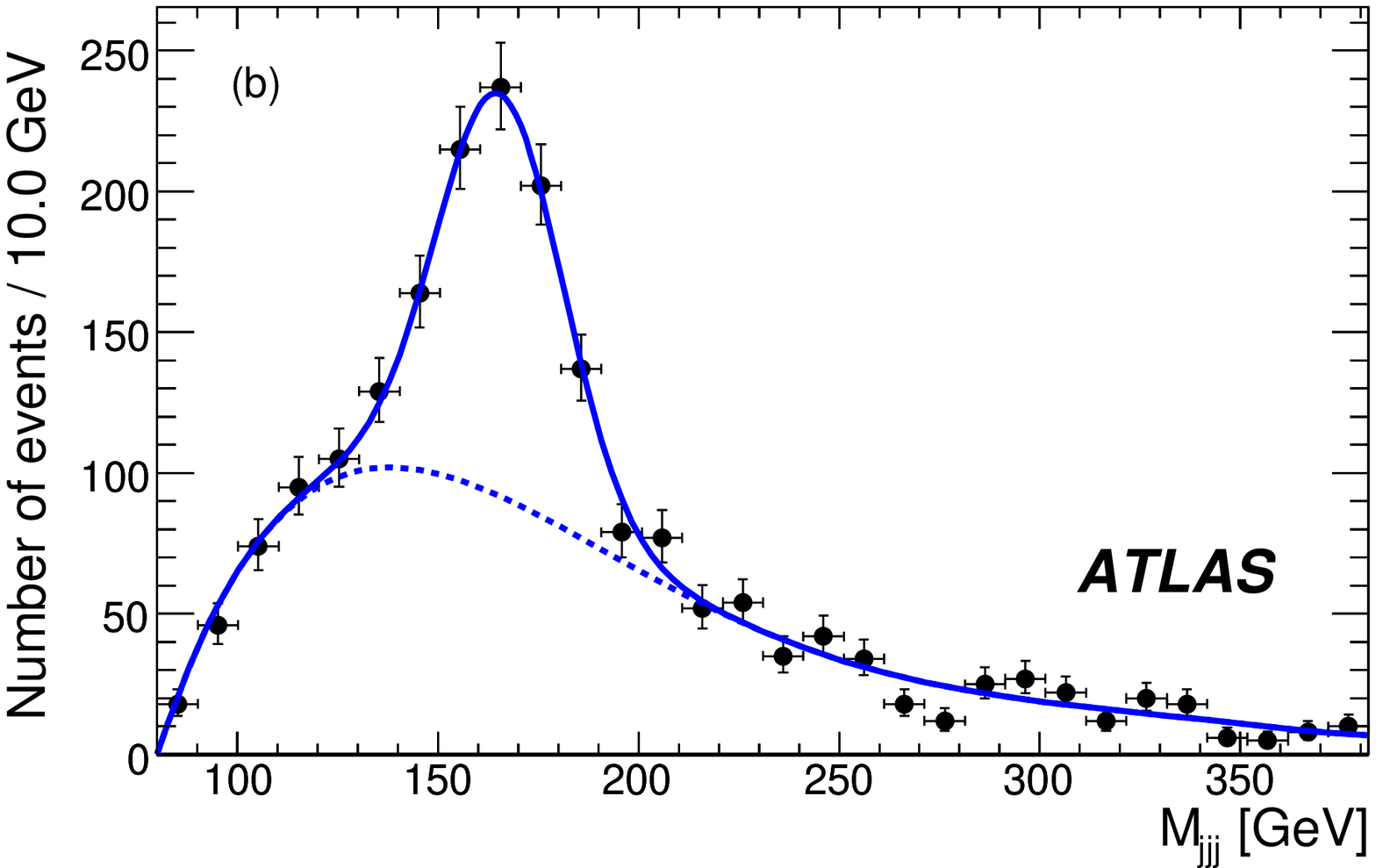}
\caption{(a) Expected invariant three jet mass distribution of correctly reconstructed $\ttbar$ events (white), incorrectly reconstructed events due to jet combinatorics (dark shaded) and background processes (light shaded). The histogram is normalised to 100pb$^{-1}$ using the total available statistics (1 fb$^{-1}$). (b) Fit of a Gaussian and Chebychev polynomial to a Monte Carlo generated pseudo-experiment corresponding to 100pb$^{-1}$ of data.} 
\label{fig:top_mass_plot}
\end{figure*}

\begin{samepage}
In the counting method the cross section is determined by subtracting the number of expected background events from the number of observed events in the $M_{jjj}$ distribution divided by the selection efficiency and integrated luminosity. The main background comes from $W$+ jets and single top. When tagging of $b$-quark jets is possible, the purity can be increased by a factor four, while the efficiency is only reduced by a factor two. The largest systematic uncertainty ($\sim$10\%) in this analysis is the background normalisation.
\end{samepage}

In the likelihood fit, only the correctly reconstructed $\ttbar$ events that end up in the peak of the distribution are used for the cross section measurement. The peak is fitted with a Gaussian, while the combined background shape from jet combinatorics and other processes such as $W$+ jets is described by a Chebychev polynomial. The cross section is calculated from the number of $\ttbar$ events in the peak divided by the overall efficiency for a $\ttbar$ event to end up in the peak and the integrated luminosity. The fit is sensitive to the shapes of the distribution and therefore this is the main systematic uncertainty ($\sim$10\%). The expected uncertainties on the cross section for 100 pb$^{-1}$ of data are:
\begin{center}
\begin{tabular}{l r r r r}
  Likelihood method     & 7 (stat) & $\pm 15$ (syst) & $\pm 3$ (PDF) $\pm 5$ (luminosity) $\%$     \\
  Counting method       & 3 (stat) & $\pm 16$ (syst) & $\pm 3$ (PDF) $\pm 5$ (luminosity) $\%$     \\
\end{tabular}
\end{center}

\section{DIFFERENTIAL CROSS SECTION}
The measurement of the differential cross section as function of the invariant mass $M_{\ttbar}$ of the $\ttbar$ system provides an important check of the Standard Model. Deviations from the $\ttbar$ continuum indicate the presence of new physics, for example new heavy resonances decaying into a $\ttbar$ pair. Semileptonic events are selected using the same criteria as mentioned in the previous section, i.e. without $b$-tagging. The $\ttbar$ pair is reconstructed by adding up the vectors of the four highest $p_{T}$ jets, the lepton and $\Et$. By using a $W$- and top mass constraint in a least square fit, assigning jets to the (anti-)top, a better result can be obtained than with the default event reconstruction only combining the four jets with the lepton and $\Et$ using a $W$-mass constraint on the leptonic side. The expected mass resolution ranges from 5\% to 9\% between 200 and 850 GeV/c$^{2}$. 

Also, the double differential cross section for $\ttbar$ is sensitive to possible new physics. In this measurement $b$-tagging is used in addition to the default single lepton $\ttbar$ selection criteria to improve purity up to 45\%. The hadronic top is reconstructed by finding the highest $p_{T}$ combination of a $b$-tagged jet with a di-jet combination $\lvert M_{jj} - M_{W}\rvert < 20$ GeV/c$^{2}$ close to it. The region which is covered by this method is 50 GeV/c $< p_{T} <$ 280 GeV/c and rapidity $\lvert y \rvert < 2$. The statistical error on the distribution will be around 30\% for 100 pb$^{-1}$ (10\% for 1 fb$^{-1}$). The main systematic uncertainty comes from the jet energy scale and initial/final state radiation ($\sim$15\%).
\begin{figure*}[ht]
\centering
\includegraphics[width=.3\textwidth]{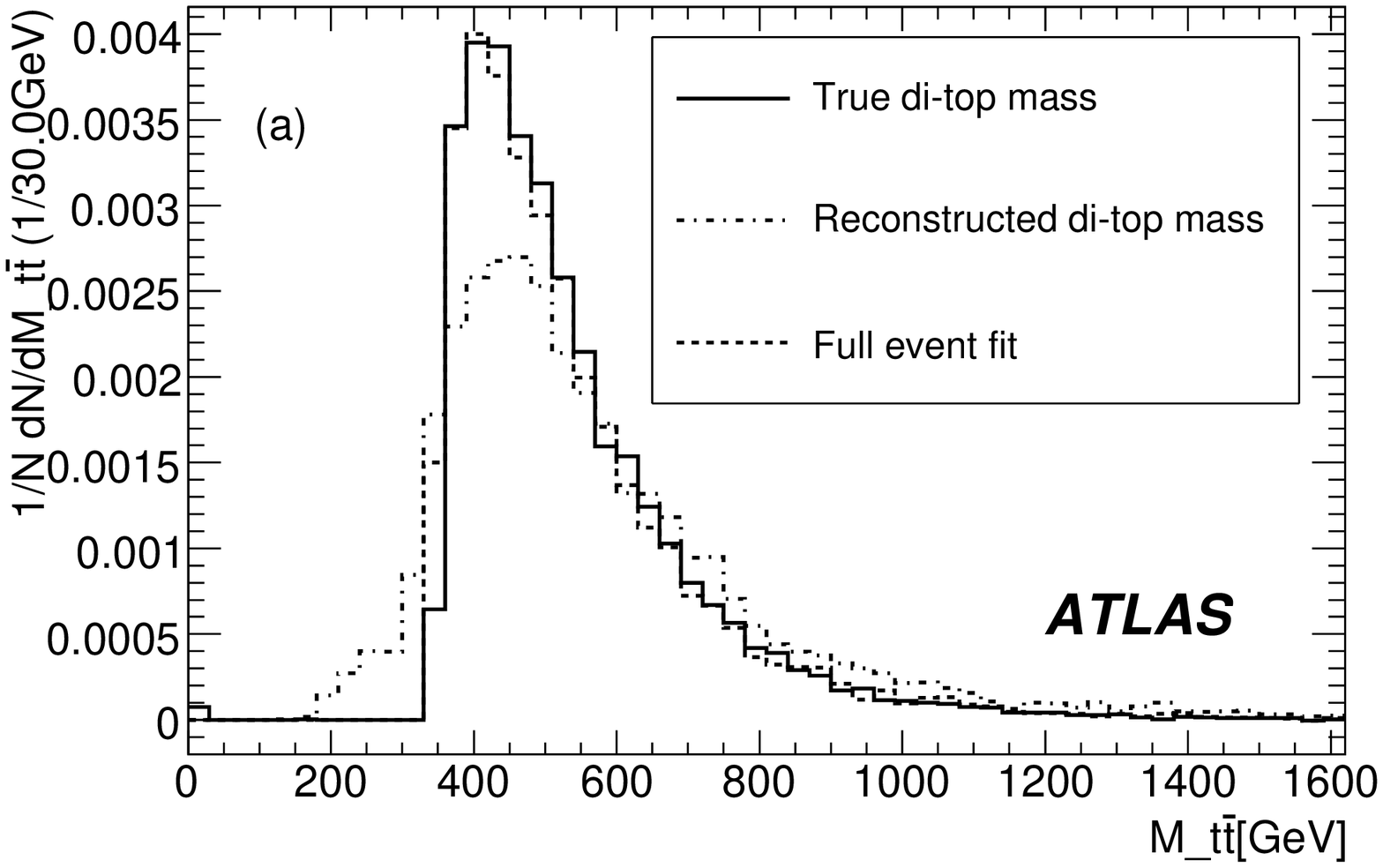}  
\hspace{.1\textwidth}
\includegraphics[width=.3\textwidth]{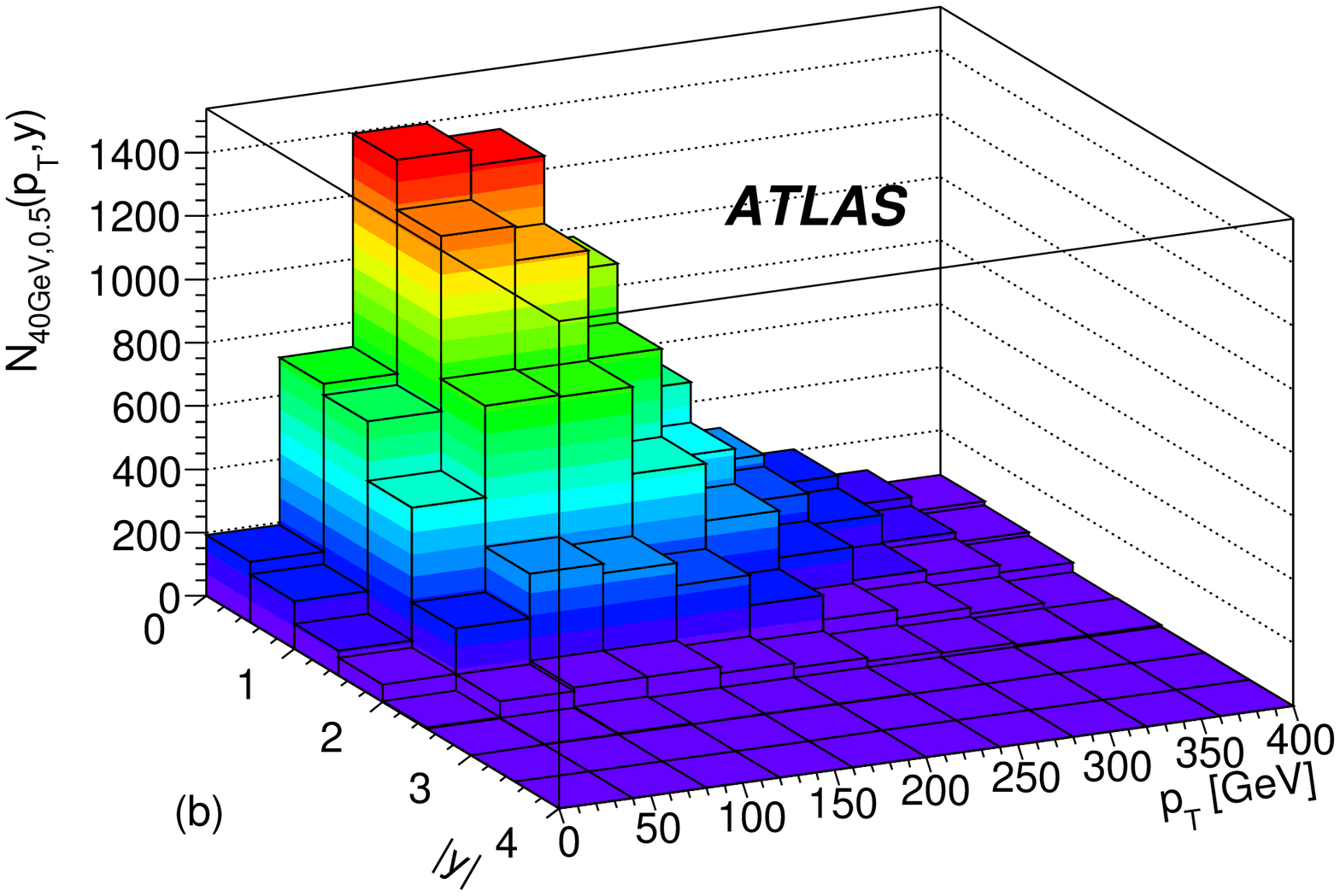}
\caption{(a) Normalised di-top mass distribution for the default (dotted line) and the improved (dashed line) reconstruction and (b) reconstructed $p_{T}$ and $y$ distribution of hadronically decaying top quarks, normalised to 1 fb$^{-1}$.} 
\label{fig:differential_xsec}
\end{figure*}

\section{DILEPTON CHANNEL}
For the cross section determination in the dilepton channel three measurements are considered. The three measurements start with the preselection of events with two high-$p_{T}$ opposite signed leptons ($e^{+}e^{-}$, $\mu^{+}\mu^{-}$ and $e^{\pm}\mu^{\mp}$) and then have different additional requirements. These events are efficiently selected by a combination of single and dilepton triggers. The uncertainty is expected to be small due to the trigger {\texttt{OR}} condition between the channels and the high statistics available for efficiency measurements such as the 'tag-and-probe' method using events with $Z$-bosons.

In the cut and count method the cross section is determined by comparing the number of observed events with the number of expected background events (from Monte Carlo). The optimum selection criteria are determined to be: two isolated opposite signed leptons with $p_{T}>20$ GeV/c, a veto on events with $M_{\ell^{+}\ell^{-}}$ around $Z$ peak (85-95 GeV/c$^{2}$), two jets of at least 20 GeV/c and $\Et>30$ GeV. With a S/B of 4.3 the main background processes remaining are $Z\rightarrow \tau^{+} \tau^{-}$ and semi-leptonic $\ttbar$ where one of the jets fakes a lepton. The jet energy scale introduces the largest systematic uncertainty.

The inclusive template method is based on the observation that the three dominant sources of isolated leptons which can be selected in the $e\mu$ channel are $\ttbar$, $W^{+}W^{-}$ and $Z\rightarrow\tau^{+}\tau^{-}$. These three processes can be separated by looking at the 2-D plane spanned by $\Et$ and $N_{jet}$. After scanning different configurations of background templates, the template with the highest probability is selected. The fit has free parameters including the $\ttbar$, $W^{+}W^{-}$ and $Z\rightarrow\tau^{+}\tau^{-}$ cross section. Contamination from processes with a fake lepton is reduced by using tight isolation criteria for the electron and a veto on events with $\Et$ aligned along the reconstructed muon. The acceptance of the two leptons and the shapes of the 2-D templates determine the systematic uncertainties.

For the third method a log-likelihood function is used to extract the parameters $N_{sig}$ and $N_{bkg}$ given the fixed total number of events $N_{tot}$. The signal and background input functions are determined by fitting Chebychev polynomials to Monte Carlo generated distributions. The sum of the semi-leptonic $\ttbar$, $W^{+}W^{-}$ and $Z\rightarrow\tau^{+}\tau^{-}$ is considered as a single background distribution. The two variables for the distributions are $\Delta\varphi$ between: (i) the highest $p_{T}$ lepton and $\Et$ and (ii) the highest $p_{T}$ jet and $\Et$. Like for the cut and count method, the jet energy scale is the dominant systematic uncertainty. Expected uncertainties on the cross section for 100 pb$^{-1}$ of data are:
\begin{center}
\begin{tabular}{l r r r r}
  Cut and Count method  & 4 (stat) & $^{+5}_{-2}$ (syst) & $\pm 2  $  (PDF) $\pm 5$ (luminosity) $\%$     \\
  Template method       & 4 (stat) & $\pm 4$      (syst) & $\pm 2  $  (PDF) $\pm 5$ (luminosity) $\%$     \\
  Likelihood method     & 5 (stat) & $^{+8}_{-5}$ (syst) & $\pm 2  $  (PDF) $\pm 5$ (luminostiy) $\%$     \\
\end{tabular} 
\end{center}

\begin{figure*}[ht]
\centering
\includegraphics[width=.3\textwidth]{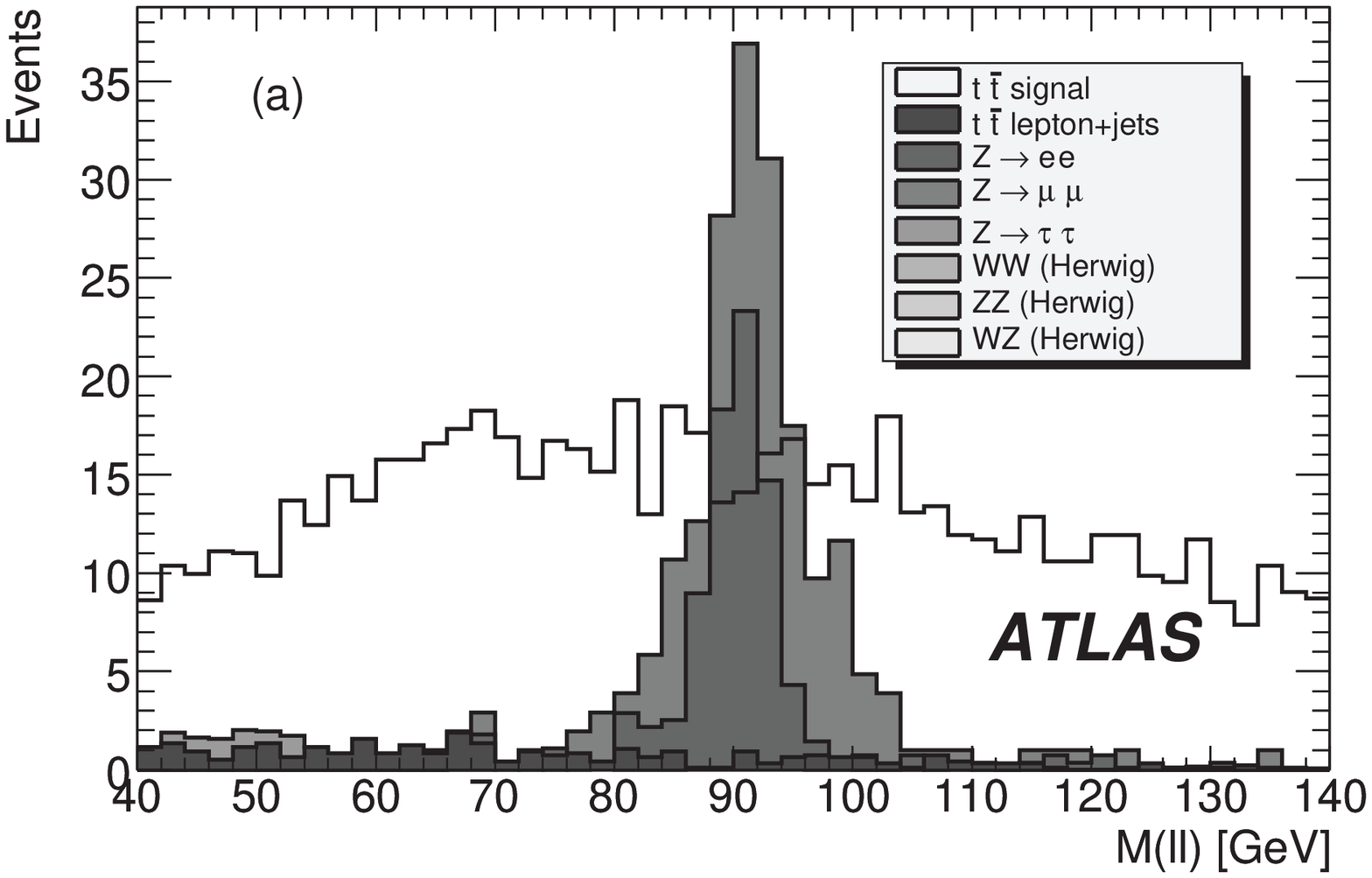}  
\includegraphics[width=.3\textwidth]{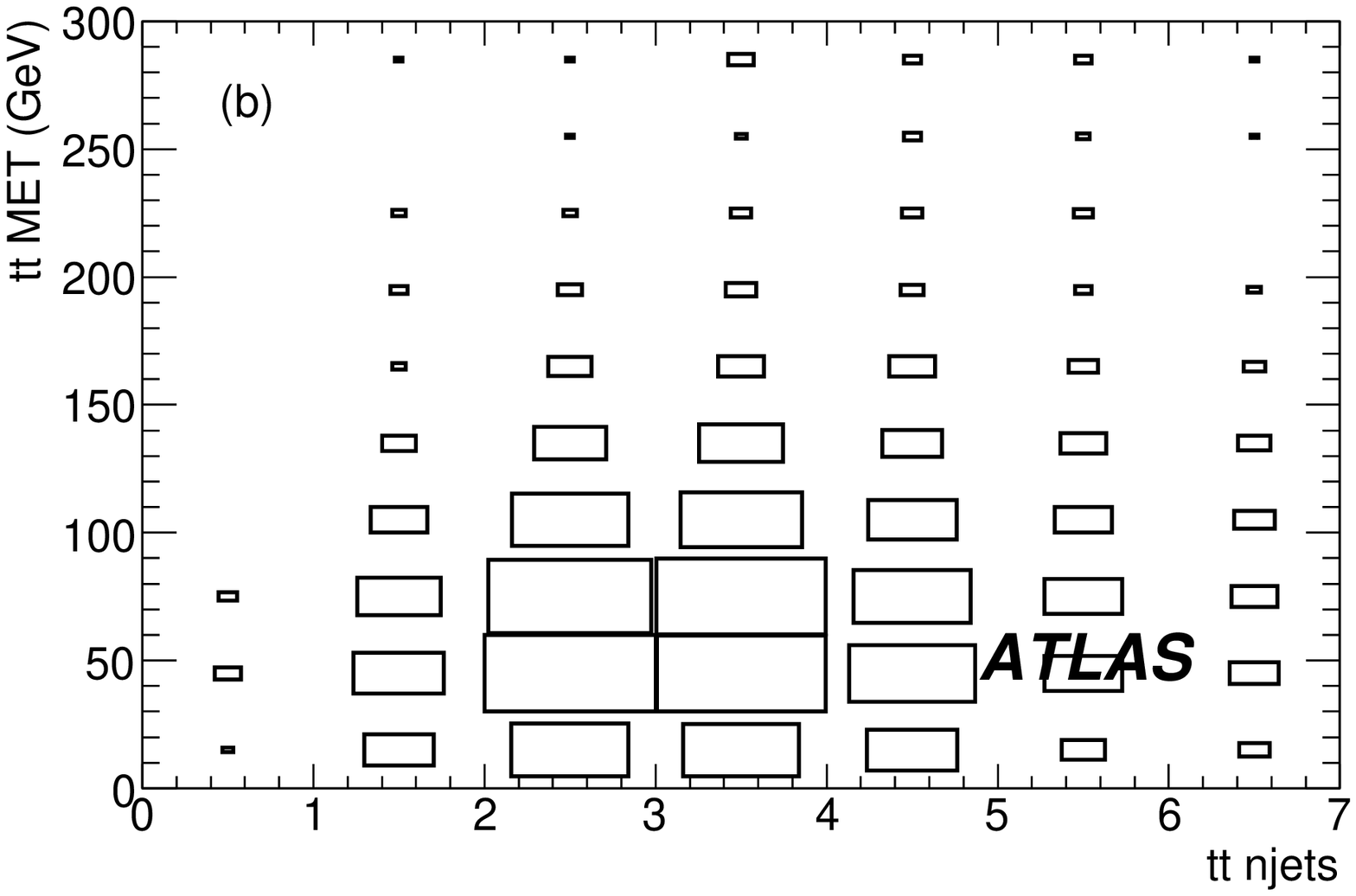}
\includegraphics[width=.3\textwidth]{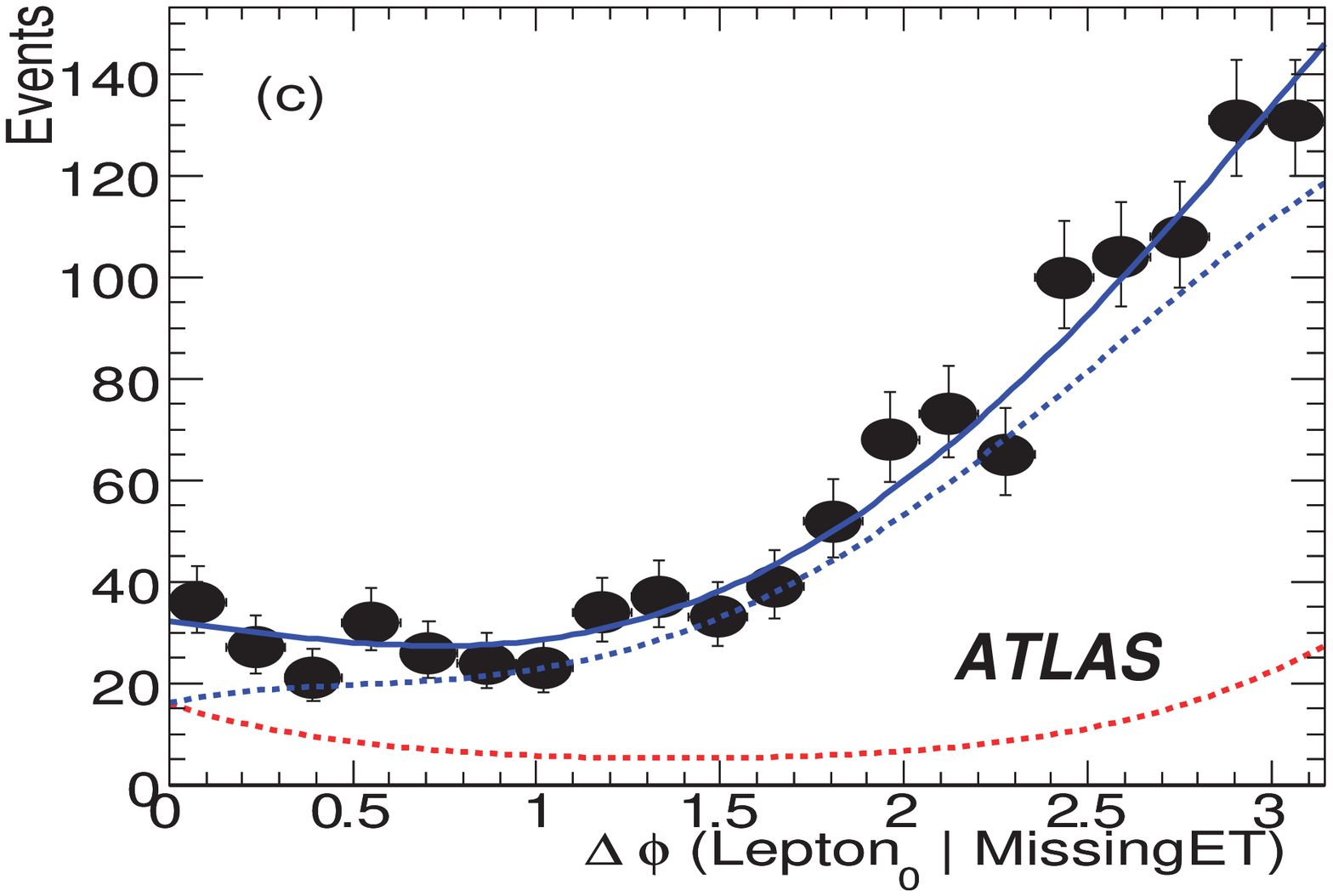}
\caption{(a) Invariant di-lepton mass distribution in the cut \& count method, (b) $\ttbar$ template of the $N_{jet}$ vs $\Et$ distribution for the inclusive template fit and (c) a likelihood fit to the signal (dotted blue line) and background (dotted red line).} 
\label{fig:dilepton_channel}
\end{figure*}

\begin{acknowledgments}
The author has been supported by the Netherlands Organisation of Scientific Research (NWO) under VIDI research grant 680.47.218.
\end{acknowledgments}

\end{document}